\documentclass[aps,prd,twocolumn,showpacs,preprintnumbers,%
  nofootinbib,amsmath,amssym,superscriptaddress]{revtex4-1}

\usepackage[T1]{fontenc}
\usepackage{color}
\usepackage[colorlinks=true,linktocpage=true,%
  linkcolor=blue,citecolor=blue,urlcolor=blue]{hyperref}
\usepackage{graphicx}
\usepackage{xcolor}
\newcommand{\calE}{\mathcal{E}}
\newcommand{\calL}{\mathcal{L}}
\newcommand{\calO}{\mathcal{O}}
\newcommand{\bj}{\boldsymbol{j}}
\newcommand{\bp}{\boldsymbol{p}}
\newcommand{\bu}{\boldsymbol{u}}
\newcommand{\bx}{\boldsymbol{x}}
\newcommand{\bz}{\boldsymbol{z}}
\newcommand{\bB}{\boldsymbol{B}}
\newcommand{\bH}{\boldsymbol{H}}
\newcommand{\bJ}{\boldsymbol{J}}
\newcommand{\bL}{\boldsymbol{L}}
\newcommand{\bM}{\boldsymbol{M}}
\newcommand{\bS}{\boldsymbol{S}}
\newcommand{\bmu}{\boldsymbol{\mu}}
\newcommand{\bomega}{\boldsymbol{\omega}}
\newcommand{\bnabla}{\boldsymbol{\nabla}}
\newcommand{\bzero}{\boldsymbol{0}}
\begin{document}
\title{Eddy magnetization from the chiral Barnett effect}
\author{Kenji Fukushima}
\affiliation{Department of Physics, The University of Tokyo, %
             7-3-1 Hongo, Bunkyo-ku, Tokyo 113-0033, Japan}
\author{Shi Pu}
\affiliation{Department of Physics, The University of Tokyo, %
             7-3-1 Hongo, Bunkyo-ku, Tokyo 113-0033, Japan}
\affiliation{Department of Modern Physics, University of Science and \\ %
Technology of China,  Hefei, Anhui 230026, China}
\author{Zebin Qiu}
\affiliation{Department of Physics, The University of Tokyo, %
             7-3-1 Hongo, Bunkyo-ku, Tokyo 113-0033, Japan}

\begin{abstract}
  We discuss the spin, the angular momentum, and the magnetic moment
  of rotating chiral fermions using a kinetic theory.  We find that,
  in addition to the chiral vortical contribution along the rotation
  axis, finite circular spin polarization is induced by the
  spin-momentum correlation of chiral fermions, which is canceled by
  a change in the orbital angular momentum.  We point out that the
  eddy magnetic moment is nonvanishing due to the $g$-factors,
  exhibiting the chiral Barnett effect.
\end{abstract}
\maketitle

%%%%%%%%%%   Introduction   %%%%%%%%%%
\section{Introduction}
\label{sec:intro}

The Barnett effect refers to the magnetization induced by mechanical
rotation of a charge neutral object~\cite{Barnett1915}.  The
Einstein--de~Haas effect is an inverse
phenomenon~\cite{einstein1915experimenteller}, that is, a finite
rotation attributed to a change in the magnetization.  We can
regard these two closely related effects as realization of
transmutation between the spin, $\bS$, and the orbital angular
momentum, $\bL$, via the $LS$ coupling.  Because of the conservation
law of the total angular momentum $\bJ$, a change in the magnetization
or $\bS$ must be compensated by a change in $\bL$.  For a historical
summary of the theory and the experiments, Ref.~\cite{Barnett1935} is
one of the most comprehensive reviews, in which the gyromagnetic
effects including not only the above-mentioned two effects but also
Maxwell's experiment and the gyromagnetic magnetization by rotating
magnetic fields are explained from a general point of view.

Before relativistic generalization of the Barnett effect, which is the
central subject studied in the current work, it would be useful to
review key equations briefly for the conventional Barnett effect.  A
finite rotation with the angular velocity vector $\bomega$ would shift
the one-particle energy by $\bomega\cdot\bJ$.  This energy shift
should be equated to the magnetic energy of $\bmu\cdot\bH_{\rm eff}$
with the magnetic moment $\bmu$ and the effective magnetic field
$\bH_{\rm eff}$ corresponding to the magnetization.  We note that the
vacuum permeability is $\mu_0=1$ in our convention of the natural
unit.  With the magnetic susceptibility, $\chi_B$, the magnetization
is given as $\bM=\chi_B\bH_{\rm eff}$, and the magnetic moment is
$\bmu=\gamma\bJ$ where $\gamma$ denotes the gyromagnetic ratio.
Combining these relations to eliminate $\bH_{\rm eff}$ and $\bmu$, we
finally get the well-known formula, i.e.,
$\bM=(\chi_B/\gamma)\bomega$.

The Barnett/Einstein--de~Haas effects have attracted attentions in
general physics fields including condensed matter physics for years.
Theoretical studies are found, for example, in the rotational states
of nanostructured magnetic
systems~\cite{PhysRevB.75.014430,PhysRevB.79.104410,PhysRevB.81.214423,Bauer2010PRB}.
It has also been pointed out in
Refs.~\cite{Bauer2010PRB,bauer2012spin,Landau:EM} that both the
Barnett and the Einstein--de~Haas effects are governed by the same
gyromagnetic tensor components which satisfy the Onsager reciprocal
relations, i.e., the gyromagnetic relation.  In experiments,
therefore, confirming one of them could be sufficient instead of
measuring both two effects for the same physical system.  Here, we lay
out several examples of experimental realization:  The
Einstein--de~Haas effect has been observed in thin film deposited on a
microcantilever~\cite{wallis2006einstein}.  There are several
proposals for experiments in an atomic gas with Bose-Einstein
condensate~\cite{PhysRevLett.96.080405,2017arXiv170105446E} and in a
ferromagnetic insulator with phonons~\cite{QiuPRL}.  In contrast to
the Einstein--de~Haas effect, the Barnett effect has been measured in
systems such as the magnetic nanostructures~\cite{bretzel2009barnett},
nuclear magnetic resonance~\cite{chudo2014observation}, paramagnetic
materials~\cite{OnoPRB}, etc.
Furthermore, circular spin-current generation has been theoretically
predicted as a result of spin-orbit interaction and the mechanical
rotation~\cite{PhysRevLett.106.076601,PhysRevB.87.180402,PhysRevB.96.020401},
which has an analoguous feature to what we are going to discuss in the
present work.  Interestingly, the theoretical predictions have been
experimentally confirmed recently, see
Refs.~\cite{Takahashi2015,PhysRevLett.119.077202,Hirohata2018}.  For
more details, interested readers can consult a recent
textbook~\cite{maekawa2017spin}.

Possible extension of the Barnett effect to systems with massless or
chiral fermions is an intriguing problem, and theoretical
investigations are demanded by recent experimental developments.
In high-energy experiments with almost massless quarks involved, the
most pertinent effort lies in the measurement of $\Lambda$ and
$\bar{\Lambda}$ global polarization conducted by the STAR
Collaboration of the Relativistic Heavy-Ion Collider
(RHIC)~\cite{STAR:2017ckg,Abelev:2007zk}.  In non-central collisions,
two nuclei collide with a huge orbital angular
momentum~\cite{Liang2005, Liang:2004xn, Gao:2007bc, Voloshin:2004ha,%
Betz:2007kg, Becattini:2007sr}, creating the ``most vortical fluid,''
and inducing a nonzero value of $\Lambda$ and $\bar{\Lambda}$ global
polarization.

Many works have been published to formulate the transfer from the
orbital angular momentum to the spin carried by hot and dense hadronic
matter.  Some examples include the microscopic spin-orbital coupling
model~\cite{Liang2005, Liang:2004xn, Gao:2007bc},
the statistical hydrodynamical
model~\cite{van1982, Zubarev1979, Becattini:2009wh, Becattini:2012tc, Becattini:2013fla, Becattini:2015nva, Hayata:2015lga},
and the quantum kinetic theory with Wigner functions~\cite{Gao2012, Fang:2016vpj, Pang:2017bjc}.
Moreover, it was proposed in Refs.~\cite{Liang2005, Liang:2004xn} that
the local polarization of $\Lambda$ and vector mesons could also be
experimentally sensitive to the net orbital angular momentum.  For
more relevant references and discussions, see the
review~\cite{Wang:2017jpl}.  Although there are theoretical
simulations for the observed polarization of hadrons, to deepen our
understanding from the fundamental level, it would be instructive
to analyze an idealized environment of noninteracting and rotating
chiral fermions, as done in this work.

In relativistic systems the chiral anomaly plays an important role for
inducing an imbalance with respect to
chirality~\cite{Adler1969a,Bell1969a}.  A clear manifestation of the
chiral anomaly can be found in a system with electromagnetic fields
and/or finite vorticity;  one can easily understand the chiral anomaly
in such a system in terms of helicity conservation;  the helicity of
fermions can be interchanged with the magnetic helicity and/or the
fluid helicity.  Then, helicity transmutation results in the
chiral magnetic effect
(CME)~\cite{Vilenkin1980a,Fukushima2008,Fukushima2010b}, the chiral
vortical effect (CVE)~\cite{Vilenkin1979a,Erdmenger2009,Son2009}, and
related topological effects (see
Refs.~\cite{Kharzeev2016,Liao2015,Burkov2015} for reviews).  These
topological effects induce nondissipative currents and survive in the
hydrodynamic limit.  Thus, the quantum anomaly could be
macroscopically manifested.  Interestingly, it has been argued within
the framework of hydrodynamics~\cite{Zakharov2015a} that the
transmutation between the helicity of fermions and the fluid helicity,
which is related to the CVE, can be regarded as an analogue of the
Barnett/Einstein--de~Haas effects.

In this paper we will first discuss the properties of rotating chiral
fermions using the kinetic theory and then address a possible connection
to the hydrodynamic counterpart.  We take this strategy since in this
way a physical interpretation of the spin and the orbital parts would
be transparent combined with field-theoretical considerations.  The
Boltzmann equation assumes a quasi-particle approximation, which is a
semiclassical treatment of dynamics.  Recently it has been established
how to implement the spin degrees of freedom in the Boltzmann
equation.  Such an augmented Boltzmann equation is commonly called the
chiral kinetic theory (CKT) in the high-energy physics community.
There are a number of literature using different ways to derive the
CKT;  e.g., effective field
theories~\cite{Son2012,Stone2013,Son2013,Chen2015},
path integrals~\cite{Stephanov2012,Chen:2013iga,Chen2014},
and Wigner functions~\cite{Chen2013, Hidaka2017,Hidaka:2017auj, Hidaka:2018ekt, Gao:2017gfq,Gao:2018wmr,Huang:2018wdl}.

Once we have the CKT, it is straightforward to derive the macroscopic
currents and the energy-momentum tensor.  By integrating physical
observables weighted with the distribution function
over the momentum, one will obtain the expectation values of the
observables such as the vector and axial vector currents with quantum
corrections, which are identified as the CME and the
CVE~\cite{Gao2012, Chen2013,Chen:2013iga, Chen2014, Hidaka2017}.
Remarkably, the correct transport coefficient of the CVE contains two
origins.  The first one comes from a shift in the particle energy
dispersion modified by the rotation.  The nontrivial Lorentz
transformations for massless particles, called the side-jump
effects~\cite{Chen2013, Chen2014, Hidaka2017,Hidaka:2017auj}, make
the second contribution to the CVE coefficient.
Since the description of the CVE in terms of the CKT has been
fully established, it is natural for us to employ the CKT for the
Barnett effect that is related to the CVE{}.

The present paper is organized as follows.  In Sec.~\ref{sec:angular}
we will give a brief review on the total angular momentum, the orbital
angular momentum, and the spin of chiral fermions.  In
Sec.~\ref{sec:kinetic} we will write down the expressions of the
orbital angular momentum and the spin operators in the kinetic theory
language.  In Sec.~\ref{sec:rotating} we will consider the CKT in a
globally rotating chiral system and will compute the orbital angular
momentum and the spin.  Next, we will relate our results to the
Einstein--de~Haas and the Barnett effects and will discuss their chiral
extensions in Sec.~\ref{sec:applications}.  We will also make a
comment on the anomalous hydrodynamics in Sec.~\ref{sec:hydro}.
Finally we summarize our results in Sec.~\ref{sec:conclusion}.
Throughout this paper we use the natural unit for the speed of light,
$c=1$, while we retain $\hbar$.

%%%%%%%%%%   Angular Momentum Decomposition   %%%%%%%%%%
\section{Angular Momentum Decomposition}
\label{sec:angular}

The angular momentum is a conserved quantity, but its decomposition
into the spin and the orbital components is not unique in relativistic
theories. In this section, we clarify our convention and explain its
physical interpretation.  Let us start with a free Dirac field (where
the generalization to include interaction is not difficult by
$\partial_\mu\to D_\mu$), whose Lagrangian density is,
\begin{equation}
  \calL = \bar{\psi} \bigl( i\hbar\gamma^\mu \partial_\mu
  - m \bigr) \psi\,.
\end{equation}
This Lagrangian is invariant under an infinitesimal rotation,
\begin{equation}
  x^\mu \;\rightarrow\; x^{\prime \mu} = x^\mu + \epsilon^\mu_{\;\nu} x^\nu\,,
  \label{eq:rot_trans}
\end{equation}
where $\epsilon^\mu_{\;\nu}$ is an antisymmetric tensor whose
magnitude is infinitesimally small.  The angular momentum tensor is
the N\"{o}ther current associated with rotation symmetry. We note
that, under Eq.~\eqref{eq:rot_trans}, the spinor transforms as
\begin{equation}
  \begin{split}
    \psi(x) \to \psi'(x') %&= \psi(x) + \epsilon_{\mu\nu}
    %\delta\psi^{\mu\nu}(x) \\
    &= \psi(x)-\frac{i}{2} \epsilon_{\mu\nu} \Sigma^{\mu\nu}\psi(x)\,,
  \end{split}
  \label{eq:rot_trans_spinor}
\end{equation}
where $\Sigma^{\mu\nu}=(i/4)[\gamma^\mu,\gamma^\nu]$.
Correspondingly, the N\"{o}ther current with current index $\lambda$
gets two contributions; the coordinate part from
Eq.~\eqref{eq:rot_trans} and the spinor part from
Eq.~\eqref{eq:rot_trans_spinor} as
\begin{equation}
  J^{\lambda\mu\nu} = L^{\lambda\mu\nu} + S^{\lambda\mu\nu}\,.
  \label{eq:JLS}
\end{equation}
We can express the first term, $L^{\lambda\,\mu\nu}$, using the
canonical energy-momentum tensor,
\begin{equation}
  T^{\mu\nu} = \frac{\partial \calL}{\partial (\partial_\mu\psi)}\;
  \frac{\partial\psi}{\partial x_\nu} = \bar{\psi}\, i\hbar \gamma^\mu
  \partial^\nu \psi\,,
\label{eq:T}
\end{equation}
as the following form,
\begin{align}
  L^{\lambda\mu\nu}
  &= x^\mu T^{\lambda\nu} - x^\nu T^{\lambda\mu}\ \notag\\
  &= \bar{\psi}\,i\hbar \bigl( \gamma^\lambda x^\mu \partial^\nu
  - \gamma^\lambda x^\nu \partial^\mu \bigr) \psi\,.
\label{eq:L}
\end{align}
For the second term of Eq.~\eqref{eq:JLS}, the explicit form reads,
\begin{equation}
  S^{\lambda\mu\nu} =
  \frac{\partial\calL}{\partial(\partial_\lambda \psi)}\;
  \Bigl[-\frac{i}{2}\Sigma^{\mu\nu}\psi(x)\Bigr] = \frac{1}{4}\bar{\psi}\,i\hbar
  \gamma^\lambda \bigl[ \gamma^\mu,\gamma^\nu \bigl] \psi\,.
  \label{spin_01}
\end{equation}
One could also obtain another form of the spin tensor from the
symmetrized Dirac Lagrangian,
$\calL = \frac{i\hbar}{2} \bigl(\bar{\psi} \gamma^\mu
\overrightarrow{\partial_\mu} \psi -\bar{\psi} \gamma^\mu
\overleftarrow{\partial_\mu} \psi \bigr)$, that is,
 $S^{\lambda \mu \nu}=
 \frac{1}{8}\bar{\psi}\,i\hbar \{ \gamma^\lambda,
 \bigl[ \gamma^\mu,\gamma^\nu \bigl] \}\psi$, but we are interested in
$S^{0\mu\nu}$ components for later discussions and the difference
 from Eq.~\eqref{spin_01} is vanishing and the choice of the
 Lagrangian is irrelevant for physical quantities as it should.

Now, the total angular momentum tensor is
\begin{equation}
  J^{\lambda\mu\nu} = \bar{\psi}\,i\hbar \Bigl( \gamma^\lambda x^\mu
  \partial^\nu - \gamma^\lambda x^\nu \partial^\mu
  + \frac{1}{4} \gamma^\lambda \bigl[\gamma^\mu, \gamma^\nu \bigl]
  \Bigr)\psi\,,
\label{eq:J}
\end{equation}
whose $\lambda=0$ component is the conserved charge density, i.e., the
conserved angular momentum.  Using the Dirac equation, we can easily
check that
\begin{equation}
  \partial_\lambda L^{\lambda\mu\nu}
  = -\partial_\lambda S^{\lambda\mu\nu} = \bar{\psi}\,i\hbar
  (\gamma^\mu \partial^\nu - \gamma^\nu \partial^\mu )\psi\,,
\label{eq:conserve}
\end{equation}
from which $\partial_\lambda J^{\lambda\mu\nu}=0$ immediately
follows.  If the surface term is irrelevant, we can then arrive at the
angular momentum conservation law:
\begin{equation}
  \frac{d}{dt}\,\int d^3 x\, J^{0\mu\nu} = 0\,.
\end{equation}
One might have thought that the above identification of $L^{0\mu\nu}$
and $S^{0\mu\nu}$ as the orbital and the spin components would be the
most natural.  Indeed, in the nonrelativistic limit, $L^{0\mu\nu}$ and
$S^{0\mu\nu}$ amount to the orbital and the spin components,
respectively.  Nevertheless, in relativistic theories, no unique
decomposition is guaranteed.

Actually, the energy-momentum tensor always has ambiguity by an
arbitrary antisymmetric tensor $\Sigma^{\mu\nu\lambda}$ as
\begin{equation}
  \Theta^{\mu\nu} = T^{\mu\nu} + \partial_\lambda
  \Sigma^{\mu\nu\lambda}\,.
\end{equation}
It is obvious that $\Theta^{\mu\nu}$ also satisfies the conservation
law, so it is equally qualified as the energy-momentum tensor.  In
particular, with an appropriate choice of $\Sigma^{\mu\nu\lambda}$,
one can make $\Theta^{\mu\nu}$ symmetric as
\begin{equation}
  \Theta^{\mu\nu} = \frac{1}{2}\bar{\psi}\,i\hbar(
  \gamma^\mu \partial^\nu + \gamma^\nu \partial^\mu) \psi\,.
  \label{eq:Theta}
\end{equation}
The corresponding ``orbital'' component of the angular momentum,
deduced from Eq.~\eqref{eq:L} with $T^{\mu\nu}$ replaced by
$\Theta^{\mu\nu}$, is
\begin{align}
  \tilde{L}^{\lambda\mu\nu}
%  & = \frac{1}{2}\bar{\psi}\,i\hbar \bigl[ \gamma^\lambda
%    (x^\mu \partial^\nu - x^\nu \partial^\mu ) +
%    (x^\mu \gamma^\nu - x^\nu \gamma^\mu ) \partial^\lambda \bigr]\psi
%    \notag\\
  & = \frac{1}{2}L^{\lambda\mu\nu}
  + \frac{1}{2}\bar{\psi}\,i\hbar \bigl[ ( x^\mu \gamma^\nu
    - x^\nu \gamma^\mu ) \partial^\lambda \bigr] \psi\,,
  \label{eq:Ltilde}
\end{align}
and the ``spin'' component is inferred from
$\tilde{S}^{\lambda\mu\nu} = J^{\lambda\mu\nu} - \tilde{L}^{\lambda\mu\nu}$.
Interestingly, using the Dirac equation again, we can prove,
\begin{equation}
  \partial_\lambda \tilde{L}^{\lambda\mu\nu}
  = \partial_\lambda \tilde{S}^{\lambda\mu\nu} = 0\,.
  \label{eq:conserve_sep}
\end{equation}
In contrast to Eq.~\eqref{eq:conserve}, the above
relation~\eqref{eq:conserve_sep} indicates that, in this construction,
the orbital and the spin components of the angular momentum are
separately conserved (see Ref.~\cite{PhysRevLett.118.114802} for a
related discussion on electron vortices), while the canonical ones,
$L^{\lambda\mu\nu}$ and $S^{\lambda\mu\nu}$ are not.  However, this
fact does not mean any superiority of $\tilde{L}^{\lambda\mu\nu}$ and
$\tilde{S}^{\lambda\mu\nu}$ because neither of them is a true symmetry
generator alone.  The situation is quite similar to the decomposition
of the optical spin and the optical orbital angular momentum.  For
free electromagnetic fields one can generally define individually
conserved spin and orbital angular momentum operator, but due to the
transversality constraint, only their combination, i.e., the total
angular momentum is the physically meaningful
quantity~\cite{0295-5075-25-7-004,doi:10.1080/09500341003654427}.

Throughout this work we adopt the canonically defined spin $S^{\lambda\mu\nu}$
and orbital angular momentum $L^{\lambda\mu\nu}$, because these are
the definitions with most natural connection to their nonrelativistic
counterparts.  Another advantage is that $S^{0ij}$, or $\bS$, turns
out to be nothing but the axial current,
\begin{align}
  S^{0ij} %= \frac{i\hbar}{2}\bar{\psi}\gamma^0\gamma^i\gamma^j\psi
  &= \epsilon^{ijk}\,\frac{\hbar}{2}\,\bar{\psi}\gamma^k \gamma_5 \psi
  = \epsilon^{ijk}\,\frac{j_5^k}{2}\,,
    \label{eq:axialj} \\
  S^{k} &\equiv \frac{1}{2} \epsilon^{ijk} S^{0ij}\,.
\end{align}
Thus, it has an interpretation evidently related to the chiral
anomaly.  Similarly we define the orbital angular momentum vector
$\bL$ as
\begin{equation}
  L^{k} \equiv \frac{1}{2} \epsilon^{ijk} L^{0ij}\,.
\end{equation}
Equation~\eqref{eq:axialj} also implies that, if the axial
current is a measurable physical observable, so will $\bS$ and
then $\bL$ be.

%%%%%%%%%%   Incarnation in Kinetic Theory   %%%%%%%%%%
%\section{Incarnation in Kinetic Theory}
\section{Transcription to Kinetic Theory}
\label{sec:kinetic}

Since we will deal with our problem in terms of kinetic theory, we
should seek for corresponding expressions for $L^{\lambda\mu\nu}$ and
$S^{\lambda\mu\nu}$ involving the distribution function,
$f(\bp,\bx,t)$.  We note that the spin and the orbital angular
momentum are the properties of matter in equilibrium unrelated to
the collisions,  once the corresponding operators for
$L^{\lambda\mu\nu}$ and $S^{\lambda\mu\nu}$ are identified.  Although
we discuss the kinetic theory transcription, we are not studying the
off-equilibrium dynamics, but we are considering the operators in
terms of the kinetic theory in this section and in terms of
hydrodynamics in Sec.~\ref{sec:hydro}.

To this end, we consider the single-particle angular momentum
tensor as done in Ref.~\cite{Chen2015, Yang:2018lew}, i.e.,
\begin{equation}
  J^{\mu\nu} %= L^{\mu\nu} + S^{\mu\nu}
  = x^\mu p^\nu - x^\nu p^\mu + S^{\mu\nu}\,,
  \label{eq:J_kin}
\end{equation}
where $p^\mu=(p=|\bp|,\,\bp)$. Comparing Eq.~\eqref{eq:J_kin}
with Eq.~\eqref{eq:J}, given the correspondence of
$i\hbar\partial^\mu \to p^\mu$, we can identify the first two terms as
the orbital part $L^{0\mu\nu}$ of our choice. Then, the last term
represents the spin part, whose specific form, according to
Ref.~\cite{Chen2015,Hidaka2017,Hidaka:2017auj}, is fixed up to a frame
vector $n_\beta$.  We choose the laboratory frame with
$n_\beta=(1,\,\bzero)$, which simplifies the concrete expression of
$S^{\mu\nu}$, leading to the following operator decomposition:
%\begin{align}
%  L^{ij} = & x^i p^j - x^j p^i \,\\
%  S^{ij} = &\hbar\lambda\,\epsilon^{ijk}
%  (\hat{p}^{k} - \hbar\lambda \epsilon^{k mn}\frac{\hat{p}^m}{p}
%  \partial_n)
%\end{align}
%\begin{align}
%  L^{ij} = x^i p^j - x^j p^i \quad &\longrightarrow \quad
%    \bL = \bx \times \bp \,,\\
%  S^{ij} = \hbar\lambda\,\epsilon^{ijk}\hat{p}^{k} \quad
%    &\longrightarrow \quad \bS = \hbar\lambda\,\hat{\bp}\,.
%\end{align}
\begin{align}
  \bL = \bx \times \bp \,,\qquad
  \bS = \hbar\lambda\,\biggl(\hat{\bp} - \hbar\lambda\,
  \frac{\hat{\bp}}{p}\times \bnabla\biggr)\,.
\label{eq:S_kin}
\end{align}
%\begin{equation}
%  S^{\mu\nu} = \hbar\lambda\,\epsilon^{\mu\nu\alpha\beta}\,
%  \frac{p_\alpha\, n_\beta}{p\cdot n}\,.
%\label{eq:S_kin_nav}
%\end{equation}
Here $\lambda$ is the helicity, i.e., $\lambda=\pm 1/2$ and
$\hat{\bp}=\bp/|\bp|$ is the unit momentum vector.
%We choose $n_\beta=(1,\,\bzero)$ in this work, which leads to
%$S^{ij}=\hbar\lambda\,\epsilon^{ijk}\hat{p}^k$ and $S^{0\nu}=0$.
%Summarizing the above, we now find,

We emphasize the importance of the second term in $\bS$ to make the
computation consistent with the CVE and the
relation~\eqref{eq:axialj}.  This additional term originates from a
gyromagnetic effect and is nothing but a familiar Rashba spin-orbit
coupling.  Another way to think of the field-theoretical origin of
this term is the current expectation value as a derivative with
respect to the vector potential.  Then, as discussed in
Ref.~\cite{Chen2014,Hidaka2017,Hidaka:2017auj}, the current reads:
%\begin{equation}
%  \bj = \hat{\bp} - \hbar\lambda\frac{\hat{\bp}}{p}\times\bnabla\,,
%\end{equation}
\begin{equation}
  \bj = \int_{\bp} \left( \hat{\bp} - \hbar\lambda\frac{\hat{\bp}}{p}\times\bnabla\ \right)f\,,
  \label{eq:vectorj}
\end{equation}
where the second term in the parentheses appears from a magnetic
dependent term, $-\lambda\hat{\bp}\cdot\bB/|\bp|$, in the energy
dispersion relation, which is eventually transcribed into the
additional term in $\bS$ as seen above.  An interesting point worth
mentioning is that $\bnabla$ is the spatial derivative and a finite
rotation would indeed induce spatial inhomogeneity.

We note that one can understand Eqs.~\eqref{eq:S_kin} and
\eqref{eq:vectorj} easily from the well-known Gordon decomposition on
the vector current with Dirac spinors at momentum $\bp$~\cite{Stone:2015bia}, i.e.,
\begin{equation}
  \hat{\bj} = \hbar\bar{\psi}\boldsymbol{\gamma}\psi
  = \frac{\hbar}{2i p}\bigl[ \psi^\dag \bnabla \psi
    - (\bnabla\psi^\dag)\psi \bigr]
  + \frac{\hbar}{2 p}\bnabla\times (\psi^\dag \boldsymbol{\Sigma} \psi)\,,
\end{equation}
where $\Sigma^k=\epsilon^{ijk}\Sigma^{ij}$.  This is the mathematical
background for Eq.~\eqref{eq:vectorj}.  Because extra $\gamma_5$ is
irrelevant for a system with either left or right handed particles
only, the argument on the vector current can be straightforwardly
translated to the axial current in Eq.~\eqref{eq:S_kin}.

It should be noted that $\bL$ and $\bS$ have the same physical unit
but $\hbar$ in $\bL$ is hidden in the momentum $\bp$ which looks
$\calO(\hbar^0)$ in a semiclassical treatment.  Such $\hbar$ counting
is consistent with our intuition that the spin is a quantum effect but
the orbital angular momentum is a macroscopic observable, while the
full consistent treatment would require the derivative expansion.

%%%%%%%%%%   Rotating chiral fermions   %%%%%%%%%%
\section{Rotating chiral fermions}
\label{sec:rotating}

In this work we study the effect of bulk rotation of chiral matter at
constant angular velocity $\bomega$ rather than a fluid with local
vorticity.  We turn electromagnetic fields off for simplicity, and if
necessary, the generalization including electromagnetic fields is
straightforward.

For an equilibrium state in the absence of rotation, the distribution
function $f$ is homogeneous in coordinate space and isotropic in
momentum space, which means that $f$ should be a function of single
particle energy $\varepsilon$, i.e.,
$f=f(\varepsilon)~$\footnote{
  According to some
  references~\cite{Chen2014,Hidaka2017,Hidaka:2017auj} our assumption
  corresponds to the ``global equilibrium'' case because our
  distribution function is independent of $n_\beta$ up to the $\hbar$
  order.  In the ``local equilibrium'' case the distribution function
  may depend on $n_\beta$ which is generally a function of spatial
  coordinates.  For more discussions on polarization effects in the
  local equilibrium case, see Ref.~\cite{Yang:2018lew}.}.
Let us
consider what would change if we introduce $\omega\neq 0$ into the
system.  For this purpose we put ourselves into a comoving frame that
rotates together with matter.  We can thereby postulate that the local
thermal equilibrium is reached after a sufficiently long time, so that
$f=f(\varepsilon_{\text{rot}})$ with $\varepsilon_{\text{rot}}$
defined in the comoving frame (which is implicitly assumed in the
implementation of finite-temperature field theory in
Ref.~\cite{Vilenkin1979a}).  We can solve a free Weyl equation in the
rotating frame to find $\varepsilon_{\text{rot}}$ as
\begin{equation}
  \varepsilon_{\text{rot}} = p - \bomega \cdot \bigl(
    \bx \times \bp + \hbar\lambda\hat{\bp} \bigr)
\label{eq:erot}
\end{equation}
using the lab-frame (nonrotating) coordinates $\bx$ and momenta
$\bp$.  We note that the energy shift in Eq.~\eqref{eq:erot} takes a
standard cranking form, $-\bomega\cdot\bJ$.  In terms of lab-frame
variables $f(\varepsilon_{\text{rot}})$ is neither homogeneous in
coordinate space nor isotropic in momentum space due to finite
rotation, thus the spin and the orbital angular momentum derived from
$f(\varepsilon_{\text{rot}})$ can be nonzero.  We begin with
calculating the spin expectation value under an assumption that
$\omega$ is small.  Up to the linear order of $\omega$ we get,
\begin{align}
  \langle\bS\rangle &= \int_{\bp}\,
  \lambda\hbar\biggl(\hat{\bp}
  -\lambda\hbar\frac{\hat{\bp}}{p}\times\bnabla\biggr)\,
  f(\varepsilon_{\text{rot}}) \notag\\
%  &= \int_{\bp}\,
%    \lambda\hbar\hat{\bp}\, f'(p)\, (-\bomega)\cdot
%    (\bx\times\bp + \hbar\lambda\hat{\bp}) \notag\\
  &\approx -\hbar \lambda(\bomega\times\bx) \int_{\bp} \frac{p}{3}\, f'(p)
    - \hbar^2 \lambda^2 \bomega \int_{\bp}\, f'(p)
\label{eq:Srot}
\end{align}
where $f'(p)= \partial{f(p)}/{\partial p}$.  It should
be mentioned that our ``expectation value'' involves only the momentum
integration, $\int_{\bp}=\int d^3p/(2\pi\hbar)^3$, but not the
coordinate integration, which is denoted later by
$\int_V = \int d^3 x$.

Here, we briefly mention the difference between setups in
Refs.~\cite{Chen2014,Chen2015} and ours.  If the rotation effects are
introduced by local vorticity vector as in
Refs.~\cite{Chen2014,Chen2015}, physical quantities can be
homogeneous.  However, to characterize the Einstein-de~Haas effect, we
implicitly assume a finite size system, for which the center of
rotation is well-defined.  Then, the velocity of rotating particles
depends on the distance from the center of rotation, and physical
quantities including the spin expectation value can be dependent on
$x$ as seen in the first term in Eq.~\eqref{eq:Srot}.

We shall make a remark about our power counting of $\hbar$ order.
In the last section we found the operators for the spin and the
orbital angular momentum in a heuristic way.  In principle, one could
utilize the Wigner function to take account of quantum corrections
systematically in the $\hbar$ expansion.  Then, $\bS$ and
$\varepsilon_{\text{rot}}$ may have $\calO(\hbar^3)$ and
$\calO(\hbar^2)$ corrections, respectively, and they contribute to an
$\calO(\hbar^3)$ correction to Eq.~\eqref{eq:Srot}.

We next turn to the orbital angular momentum.  In the same way
we expand the distribution function with respect to $\bomega$ and obtain
\begin{align}
  \langle \bL \rangle   & \approx \int_{\bp}\, (\bx\times\bp)\,f'(p)
    (-\bomega) \cdot (\bx\times\bp + \hbar\lambda\hat{\bp}) \notag \\
  &= -\bx \times (\bomega\times\bx) \int_{\bp}\, \frac{p^2}{3}\,f'(p)
  + \hbar\lambda (\bomega\times\bx) \int_{\bp}\, \frac{p}{3}\,f'(p)\,.
\label{eq:Lrot}
\end{align}
Equations~\eqref{eq:Srot} and \eqref{eq:Lrot} are our central results
in this paper.  In the following sections we shall expound their
physical interpretations.

%%%%%%%%%%   Applications   %%%%%%%%%%
\section{Applications -- chiral Einstein--de~Haas / Barnett effects}
\label{sec:applications}

We utilize our results for $\langle\bL\rangle$ and $\langle\bS\rangle$
to discuss the relativistic extension of the Einstein--de~Haas effect
and the Barnett effect.

%%%%%   Chiral EdH Effect   %%%%%
\subsection{Chiral Einstein--de~Haas Effect}
\label{sec:EdH}

\begin{figure}
\includegraphics[width=0.9\columnwidth]{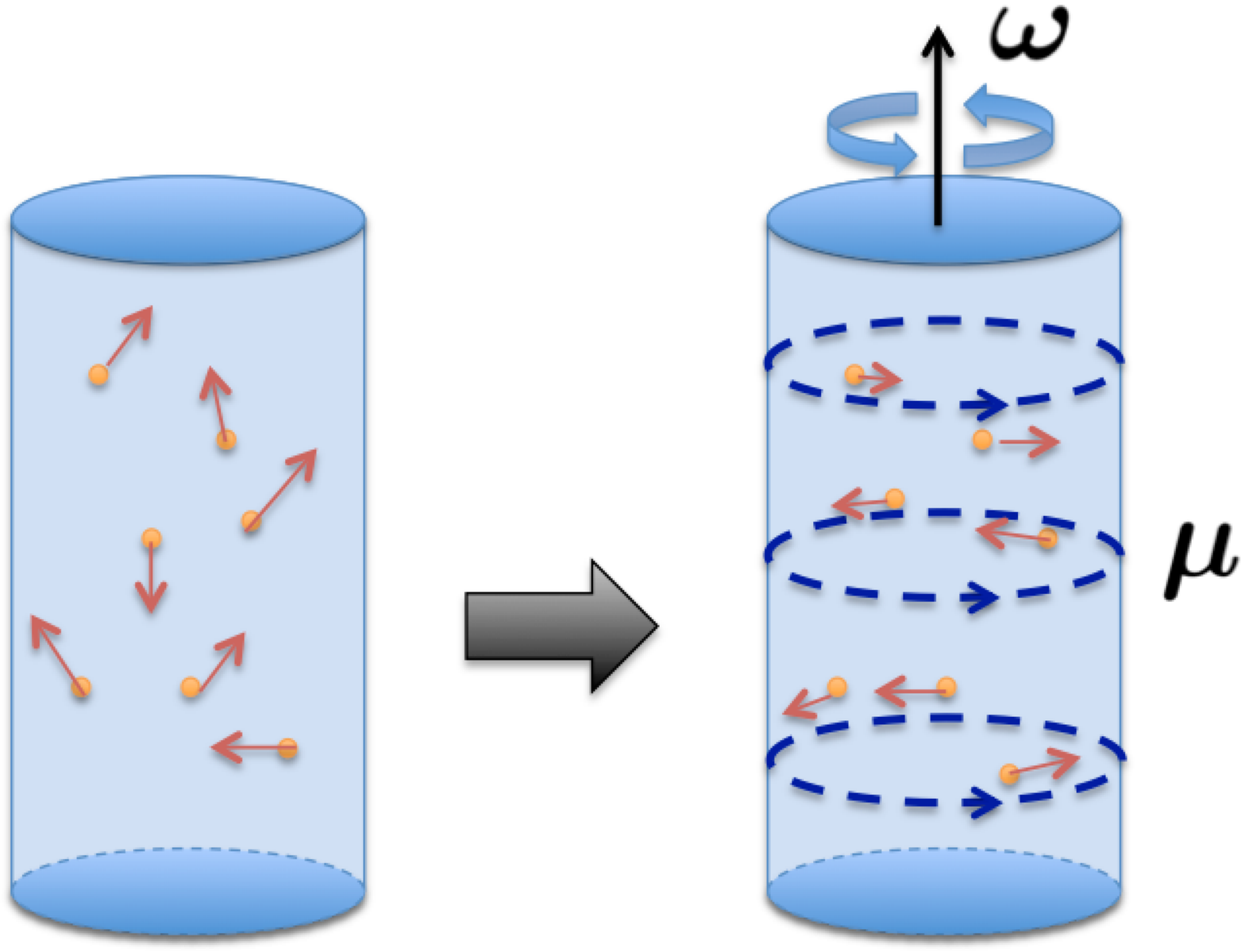}
\caption{A schematic illustration for an intuitive picture to
  understand the circular spin polarization and the associated eddy
  magnetization $\bmu$ in a rotating chiral system with the angular
  velocity vector $\bomega$.  For simplicity we only consider the
  right-handed fermions in the illustration.  The red arrows stand for
  the direction of particle momentum and spin.}
\label{fig:CBE}
\end{figure}

The physical meaning of Eq.~\eqref{eq:Srot} becomes transparent
once we add up both left-handed and right-handed contributions.
After an integration by parts, the first term in Eq.~\eqref{eq:Srot},
which is denoted by $\langle\bS\rangle_\perp$ hereafter, takes the
following form as
\begin{align}
  \langle\bS\rangle_\perp &= -\hbar\sum_{R,L} \lambda (\bomega\times
  \bx) \int_{\bp}\, \frac{p}{3}\, f'_\lambda(p) \notag\\
  &= \frac{\hbar}{2} (\bomega\times\bx)
  \int_{\bp} \bigl[ f_{R}(p)-f_{L}(p)\bigr]
  = \frac{\hbar}{2} (\bomega\times\bx)\, n_5\,.
  \label{eq:Strans}
\end{align}
where $f_{R}$ and $f_{L}$ refer to the distribution functions of
right-handed and left-handed particles respectively, and thus
$n_5=n_R-n_L$ means the chirality density.  Such a rotation-induced
spin alignment is intuitively understood as follows.  For massless
fermions, the spin and the momentum directions are locked up.
In this way, the angular momentum is related to the translational motion.
Therefore, if we macroscopically move our chiral matter with the
velocity $\bu=\bomega\times\bx$, the spin will be tilted along $\bu$.
In this sense $\langle\bS\rangle_\perp$ is a unique result inherently
for chiral fermions.  Interestingly, this transverse eddy spin
alignment requires a finite chiral imbalance.
We present a schematic illustration in Fig.~\ref{fig:CBE} to explain
how $\langle\bS\rangle_\perp$ appears in a rotating chiral system.

%---   figure   ---%
%---   figure   ---%

After similar algebra, we rewrite the orbital angular momentum
Eq.~\eqref{eq:Lrot} as
\begin{align}
  \langle \bL \rangle &= \bx \times (\bomega\times\bx) \frac{4}{3}
  \int_{\bp} p\bigl[ f_{R}(p)+f_{L}(p) \bigr] - \langle\bS\rangle_\perp
  \notag\\
  &= \langle \bL \rangle_{\text{mech}} - \langle \bS \rangle_\perp\,.
  \label{eq:Lmech-S}
\end{align}
Here, $\langle\bL\rangle_{\text{mech}}$ represents the first term
involving $\bomega\times\bx$ in the above expression.  We shall
illuminate the physical interpretation of
$\langle\bL\rangle_{\text{mech}}$ in what follows.  For concreteness
we will consider a cylindrically symmetric system which rotates
rigidly around the $z$-axis, i.e., $\bomega=\omega\hat{\bz}$.  Then in
such a setup the volume integration of
$\langle\bL\rangle_{\text{mech}}$ yields
\begin{equation}
  \int_V \langle \bL \rangle_{\text{mech}} = \omega \hat{\bz}
  \int_V\, r^2 \, \frac{4}{3}\int_{\bp}\, p
  \bigl[ f_{R}(p) + f_{L}(p) \bigr]\,.
  \label{eq:Lmech}
\end{equation}
Since $p$ is the energy for chiral fermions, the $p$ integration gives
the energy density or the mass distribution, together with which the
volume integration leads to the moment of inertia.  To see this
clearly, let us assume that the distribution function features Fermi
degeneracy to a chemical potential $\mu$, and then the energy density,
$\calE$, is calculated as $\calE=\frac{3}{4}\mu n$ where $n$ is the
number density.  Consequently,
$\frac{4}{3}\int_{\bp}\, p \bigl[ f_{R}(p) + f_{L}(p) \bigr]$ reduces
to a relativistic counterpart of the mass density,
$\mu_R n_R + \mu_L n_L$.  From this argument it is clear that
$\langle\bL\rangle_{\text{mech}}$ corresponds to the mechanically
induced orbital angular momentum, which is naturally of
$\calO(\hbar^0)$.

Next, we delve into the second term in $\langle\bL\rangle$ given by
$-\langle\bS\rangle_\perp$.  This term has an intriguing
interpretation as the ``Chiral Einstein--de~Haas effect.''
Let us consider the following thinking experiment; we rotate the
fermionic system from the initial condition,
$\langle\bL\rangle=\langle\bS\rangle=0$. Apparently, the total angular
momentum carried by rotating chiral matter should be
$\langle\bJ\rangle=\langle\bL\rangle_{\text{mech}}$.  However, as
mentioned above, due to the spin and momentum lock-up, the transverse
motion results in $\langle\bS\rangle_\perp\neq 0$.  This nonzero
$\langle\bS\rangle_\perp$ must be
canceled by a change in the orbital part so that the total angular
momentum conservation can be satisfied.  In this way, a shift by
$-\langle\bS\rangle_\perp$ should arise in $\langle\bL\rangle$.
%transferred from the spin.
Such a physical mechanism is comparable to
the Einstein--de~Haas effect.  In the nonrelativistic case the spin is
controlled by an external magnetic field, but it can be changed by the
momentum direction for chiral fermions, which induces an orbital
rotation.

We make two comments on the second term in Eq.~\eqref{eq:Srot}.  The
first one is that this term corresponding to the CVE can be also
exactly canceled in a finite size system by surface states not to
violate the angular momentum conservation~\cite{Mameda}.  The second
comment is that, if we consider the zero $n_5$ limit, the second term in
Eq.~\eqref{eq:Srot} would dominate and lead to the local spin
polarization proposed in Ref.~\cite{Gao2012}.

%%%%%   Chiral Barnett Effect   %%%%%
\subsection{Chiral Barnett Effect}
\label{sec:barnett}

Along similar lines, we can apply our formula to address
the Barnett effect for chiral fermions.  That is, a finite
magnetization is generated by rotation~\cite{Barnett1915}, which can
be quantified with our results.  For this purpose we need the
gyromagnetic ratio to convert the angular momentum into the magnetic
moment.  For nonrelativistic fermions, the gyromagnetic ratio is
derived from the Dirac equation as
\begin{equation}
  \bmu = \bmu_L + \bmu_S = g_L\,\frac{q_e}{2m}\bL
  + g_S\,\frac{q_e}{2m}\bS\,,
  \label{eq:mu_nonrela}
\end{equation}
where $q_e$ and $m$ are, respectively, the electric charge and the
mass of the considered particle.  For noninteracting Dirac fermions
the $g$-factors are $g_L=1$ and $g_S=2$.  Since $g_L\neq g_S$, the
right-hand side of Eq.~\eqref{eq:mu_nonrela} is not parallel to
$\bJ=\bL+\bS$.  Once one takes an expectation value with the $\bJ^2$
and $J_z$ eigenstates, however, one can show that the right-hand side
is projected onto the $\bJ$ direction, which is guaranteed by the
Wigner-Eckardt theorem, and the effective $g$-factor becomes the
Land\'{e} $g$-factor.

For chiral fermions Eq.~\eqref{eq:mu_nonrela} should be modified.
In the chiral limit Eq.~\eqref{eq:mu_nonrela} turns into (see
Ref.~\cite{Kharzeev2017,Stone:2015bia})%
\footnote{Infrared singularity in Eq.~\eqref{eq:mu_rela} is
  regularized by the Debye screening in many-body system.  In other
  word, the momentum convoluted with a distribution function has an
  infrared cutoff by $gT$, where $g$ is the coupling constant of the
  theory and $T$ is the temperature.}
\begin{equation}
  \bmu = \bmu_L + \bmu_S = g_L\,\frac{q_e}{2p}\bL
  + g_S\,\frac{q_e}{2p}\bS\,.
  \label{eq:mu_rela}
\end{equation}
The $g$-factors remain the same, and from now on we plug $g_L=1$ and
$g_S=2$ into $\bmu_L$ and $\bmu_S$.  We note that
Eq.~\eqref{eq:mu_rela} is a local relation, and so we compute the
expectation value as we did in the previous sections.  The results up
to $\hbar$ order are
\begin{align}
  \langle\bmu_L\rangle &= -\frac{q_e}{6}
  \bx\times(\bomega\times\bx)\int_{\bp}\, p\,f'(p) \notag\\
  &\qquad\qquad\qquad + \hbar\lambda \frac{q_e}{6}
  (\bomega\times\bx)\int_{\bp}\, f'(p)\,, \\
  \langle\bmu_S\rangle &= -\hbar\lambda\frac{q_e}{3}(\bomega\times\bx)
  \int_{\bp}\, f'(p)\,.
\end{align}
We can immediately identify the first term of $\langle\bmu_L\rangle$
as the mechanical contribution.  The integration by parts makes it
more visible as
\begin{equation}
  \langle\bmu_L\rangle_{\text{mech}} = \frac{1}{2}\bx\times
  (\bomega\times\bx)\, n_e\,,
\end{equation}
where $n_e$ represents the electric charge density.  Given that
$\bomega\times\bx$ is the velocity vector associated with the rotating
motion, the above expression is exactly the one known as the magnetic
dipole moment from the Amp\'{e}re loop in classical electromagnetism.

The second term of $\langle\bmu_L\rangle$ is at the same order as
$\langle\bmu_S\rangle$, but they do not cancel out.  The total
magnetization reads:
\begin{equation}
  \langle\bmu\rangle = \langle\bmu_L\rangle + \langle\bmu_S\rangle
  = \langle\bmu_L\rangle_{\text{mech}}
  - \hbar\lambda\frac{q_e}{6}(\bomega\times\bx)\int_{\bp}\, f'(p)\,.
  %- \hbar^2\frac{q_e}{3}\bomega\int_{\bp}\,\frac{f'(p)}{p}\,.
  \label{eq:mu}
\end{equation}
This result exhibits the chiral Barnett effect.  What is nontrivial in
the relativistic case is the second term.  It is proportional to
$\bomega\times\bx$, and thus has the circular orientation around the
rotation axis, just like previously discussed
$\langle\bS\rangle_\perp$ as shown in Fig.~\ref{fig:CBE}.  Since the
magnetic moment is a source for the magnetic field, we can expect a
generation of eddy magnetic field in rotating chiral media.  Further
exploration on this point and applications to astrophysical objects
will be reported elsewhere~\cite{Qiu}.

%%%%%%%%%%   Hydrodynamics   %%%%%%%%%%
\section{Comments on Hydrodynamical Formulation}
\label{sec:hydro}

In this section we briefly address the problem of calculating the
orbital angular momentum in anomalous hydrodynamics~\cite{Son:2009tf}.
In the framework of anomalous hydrodynamics, the energy-momentum
tensor reads (see Refs.~\cite{Gao2012, Pu:2010as}):
\begin{equation}
  T_{\text{hydro}}^{\mu\nu} = (E+P)u^\mu u^\nu - P\,g^{\mu\nu}
  + \hbar\, n_5 (u^\mu \omega^\nu + u^\nu \omega^\mu)\,,
  \label{eq:Thdyro}
\end{equation}
where $E$ and $P$ are the energy density and the pressure
respectively, and $u^\mu=\gamma(1,\,\bu)$ denotes the fluid velocity.
For simplicity we assume the small velocity limit, i.e.,
$|\bu|\ll 1$ and $u^\mu\approx(1,\,\bu)$.  In such limit the vorticity
is
$\omega^\nu=\frac{1}{2}\epsilon^{\nu\alpha\beta\gamma}
u_\alpha \partial_\beta u_\gamma \approx (0,\,\bnabla\times\bu) + \calO(|\bu|^2)$ by definition.
Using this energy-momentum tensor to define the hydro orbital angular
momentum, we find,
\begin{align}
  L_{\text{hydro}}^{ij}
  &= x^i T_{\text{hydro}}^{0j} - x^j T_{\text{hydro}}^{0i} \notag\\
  &= x^i [(E+P)u^j + \hbar\, n_5\omega^j] - (i\leftrightarrow j)\,.
\end{align}
For a mechanically rotating fluid, we specify
$u^\mu=(1,\bomega\times\bx)$ and $\omega^\mu=(0,\bomega)$, which leads
to
\begin{equation}
  \bL_{\text{hydro}} = (E+P)(\bx\times\bu)
  - \hbar\, n_5 (\bomega\times\bx)\,.
  \label{eq:Lhydro}
\end{equation}
To see the connection between Eq.~\eqref{eq:Lhydro} and our results in
a kinetic picture, we remind that massless noninteracting systems has
the equation of state as $P=E/3$, and in a kinetic framework $E$ can
be expressed as
\begin{equation}
  E = \int_{\bp}\, \frac{(u\cdot p)^2}{p^0} \,(f_{R} + f_{L})\,\approx \int_{\bp}\, p \,(f_{R} + f_{L}).
\end{equation}
Then, eventually, the hydro angular momentum takes the form of
\begin{equation}
  \bL_{\text{hydro}} = \bx\times(\bomega\times\bx) \; \frac{4}{3}
  \int_{\bp} p (f_{R}+f_{L}) - 2\langle \bS\rangle_\perp\,.
  \label{eq:Lhydro_kin}
\end{equation}
Now we can make a direct comparison between Eq.~\eqref{eq:Lhydro_kin}
and our results in Eq.~\eqref{eq:Lmech-S}.
We find that the first term corresponding to
$\langle\bmu_L\rangle_{\text{mech}}$ exactly agrees, but the
coefficient for $\langle\bS\rangle_\perp$ is different.

This discrepancy originates from different definitions of the
energy-momentum tensor.  The energy-momentum tensor operator
$T^{\mu\nu}$ obtained from the N\"{o}ther theorem in Eq.~\eqref{eq:T}
is not symmetric and does not correspond to the hydrodynamic
energy-momentum tensor in Eq.~\eqref{eq:Thdyro}.  In fact, it is
the symmetrized energy-momentum tensor operator $\Theta^{\mu\nu}$ in
Eq.~\eqref{eq:Theta} that corresponds to Eq.~\eqref{eq:Thdyro}.
Such symmetrized definition is prevalently adopted in anomalous
hydrodynamics (see, Refs.~\cite{Chen2015, Hidaka2017,Hidaka:2017auj}
for examples).
Accordingly, for the orbital angular momentum, $\bL_{\text{hydro}}$
corresponds to $\tilde{L}^{0\mu\nu}$ in Eq.~\eqref{eq:Ltilde} rather
than the canonical one $L^{0\mu\nu}$ in Eq.~\eqref{eq:L} used for our
CKT computation.

Here, we make a comment on the approximate spin conservation law.
Under such circumstance as massless fermions and vanishing
electromagnetic fields, the axial current $j_5^\mu$ is conserved, and
so does the spin.  In hydrodynamics one can show that this
conservation law of $\langle\bS\rangle$ follows from the expansions
with respect to $\hbar$ and $\omega$.  In terms of the fluid velocity
$u^\mu$ and the vorticity $\omega^\mu$, the conservation law of
$j_5^\mu$ is~\cite{Son2009, Gao2012}
\begin{equation}
  \partial_\mu j_5^\mu = \partial_\mu \bigl( n_5 u^\mu + \hbar\,
  \xi_5\, \omega^\mu \bigr) = 0 ,
  \label{eq:axial_conserve}
\end{equation}
where $\xi_5$ is the CVE coefficient as a function of the temperature,
the chemical potentials, etc.  For a slowly rotating fluid with
$u^\mu\approx(1,\bomega\times\bx)$ Eq.~\eqref{eq:axial_conserve} tells
us that
\begin{equation}
  \frac{d}{dt} n_5 = -\bnabla\cdot\bigl(n_5 \bomega\times\bx\bigr)
  + \hbar\,\partial_\mu(\xi_5\omega^\mu)\,.
\end{equation}
Then, for small $\omega$, it is easy to verify that
\begin{equation}
  \frac{d}{dt}\int_V \langle \bS \rangle = \frac{\hbar}{2}\int_V \frac{d}{dt} (n_5\bomega\times\bx) + O(\hbar^2) = \calO(\hbar^2,\, \omega^2).
\end{equation}
We thus conclude that, under the approximation to drop terms of
$\calO(\omega^2)$ and/or $\calO(\hbar^2)$, both our
$\langle\bL\rangle$ in the canonical definition and
$\langle \bL\rangle_{\text{hydro}}$ are equally qualified as the
orbital angular momentum.  It should be noted that in the above
argument we implicitly assumed that $x$ is of order of the unity.
This means that the system size must not be as large as $1/\hbar$ or
$1/\omega$; otherwise, the surface state is not
negligible~\cite{Mameda}.

%%%%%%%%%%   Conclusions   %%%%%%%%%%
\section{Conclusion}
\label{sec:conclusion}

In this work we systematically discussed the spin, the angular
momentum, and the magnetic momentum for rotating chiral fermions using
the framework of the CKT.
First, we gave a brief review of deriving the angular momentum tensors
as N\"{o}ther's currents.  Although the decomposition into the
orbital and the spin components is not unique, we adopted the
canonical definition in which the spin is directly related to the
axial current.
Next, we considered a globally rotating chiral system.  Combining the
contributions from a shift in the particle energy dispersion and an
extra spin-orbital coupling term in the CKT, we identified the
expectation values of the spin [see $\langle\bS\rangle$ in
  Eq.~\eqref{eq:Srot}] and the orbital angular momentum [see
  $\langle\bL\rangle$ in Eq.~\eqref{eq:Lrot}].

Based on these two expressions for $\langle\bS\rangle$ and
$\langle\bL\rangle$, we developed a physical picture of the
relativistic extension of the Einstein--de~Haas effect.  Up to
$\calO(\hbar)$ terms, the circular spin alignment is induced by the
mechanical rotation as illustrated in Fig.~\ref{fig:CBE}, which can be
intuitively understood through the fact that the rotation is
accompanied by the axial current.  Then, a shift in $\langle\bL\rangle$
is caused by $\langle\bS\rangle$ to maintain the conservation law of
the total angular momentum, which can be regarded as a relativistic
counterpart of the Einstein--de~Haas effect realized in a chiral
medium.

Furthermore, we applied our results to address the Barnett effect for
chiral fermions.  We computed the magnetic moments,
$\langle\bmu_L\rangle$ and $\langle\bmu_S\rangle$, proportional to the
orbital angular momentum and the spin, respectively.  The leading
order term in $\langle\bmu_L\rangle$ of $\calO(\hbar^0)$
%$\langle\bmu_L\rangle_{\text{mech}}$,
is exactly the one from the magnetic dipole moment obtained in
classical electromagnetism.  The next order terms in
$\langle\bmu_L\rangle$ and $\langle\bmu_S\rangle$ of $\calO(\hbar)$
will not cancel out and this nonvanishing magnetic moment exhibits
what we call the chiral Barnett effect.

Before closing our discussions in the end, we supplemented some
discussions on the anomalous hydrodynamics.  We pointed out that the
symmetric energy-momentum tensor adopted in the anomalous
hydrodynamics does not correspond to the canonical one derived from
the N\"{o}ther theorem.  Using the hydrodynamical energy-momentum
tensor, we could define another form of the orbital angular momentum
$\bL_{\text{hydro}}$, which is approximately an conserved quantity of
$\calO(\hbar)$.

There will be many possible extensions and applications of our work.
As discussed in Sec.~\ref{sec:barnett} the rotation may induce the
eddy magnetic fields, which could explain the internal structures of
the neutron star~\cite{Makishima:2014dua}
In astrophysics, generally, the magnetic field and
the rotation are commonly found in macroscopic objects, and the
chiral Barnett effect may play an intriguing role~\cite{Qiu}.  Another
interesting direction lies in possible generalization to
nonequilibrium situation.  In this study we assumed only a near
equilibrium distribution function to compute $\langle\bS\rangle$ and
$\langle\bL\rangle$ in a steady state.  The full real-time evolution
of $\langle\bS\rangle$ and $\langle\bL\rangle$ starting with some
initial condition would be a challenging future problem.

In the future approximations made in the present work should be
relaxed.  Our treatment of fermions is limited to the massless case
only, and the inclusion of finite mass effects would be a crucial step
toward phenomenological applications to relativistic heavy-ion
collision experiments.  In this work, we neglected surface terms, and
we should amend this with finite size effects taken into account.  So
far, our analysis is only up to $\calO(\hbar)$ apart from the CVE
term.  Thus, we have not included higher order nontrivial effects,
e.g., the local spin polarization effect ~\cite{Gao2012}.
We are currently making progresses to incorporate these effects
dropped in the present work.

%%%%%%%%%%   Acknowledgments   %%%%%%%%%%
\begin{acknowledgments}
  We thank
  Yoshimasa~Hidaka,
  Xu-Guang~Huang,
  Di-Lun~Yang,
  and Qun~ Wang
  for helpful discussions.
  K.~F.\ was supported by JSPS KAKENHI Grant
  No.\ 18H01211.
  S.~P.\ was supported by JSPS
  post-doctoral fellowship for foreign researchers.

\end{acknowledgments}

\bibliography{barnett}{}
\bibliographystyle{apsrev4-1}

\end{document}